# A Vision to Enhance Trust Requirements for Peer Support Systems by Revisiting Trust Theories

Yasaman Gheidar
*Telfer School of Management*
*University of Ottawa*
Ottawa, Canada
Yasaman.Gheidar@uottawa.ca

Lysanne Lessard
*Telfer School of Management*
*University of Ottawa*
Ottawa, Canada
0000-0003-3013-0044

Yao Yao
*Telfer School of Management*
*University of Ottawa*
Ottawa, Canada
yao@telfer.uottawa.ca

*Abstract*—This vision paper focuses on the mental health crisis impacting healthcare workers (HCWs), which exacerbated by the COVID-19 pandemic, leads to increased stress and psychological issues like burnout. Peer Support Programs (PSP) are a recognized intervention for mitigating these issues. These programs are increasingly being delivered virtually through Peer Support Systems (PSS) for increased convenience and accessibility. However, HCWs' perception of these systems results in fear of information sharing, perceived lack of safety, and low participation rate, which challenges these systems' ability to achieve their goals. In line with the rich body of research on the requirements and properties of trustworthy systems, we posit that increasing HCWs' trust in PSS could address these challenges. However, extant research focuses on objectively defined trustworthiness rather than perceptual trust because trustworthy requirements are viewed as more controllable and easier to operationalize. This study proposes a novel approach to elicit perceptual trust requirements by proposing a trust framework anchored in recognized trust theories from different disciplines that unpacks trust into its recognized types and their antecedents. This approach allows the identification of trust requirements beyond those already proposed for trustworthy systems, providing a strong foundation for improving the effectiveness of PSS for HCWs.

## I. Introduction

The mental health and well-being issues being faced by healthcare workers (HCWs) are known to significantly impact the quality of healthcare delivery worldwide [1], [2]. The COVID-19 pandemic has intensified the global healthcare workforce crisis, leading to increased stress and a broad spectrum of psychological issues among HCWs, including burnout, depression, anxiety, and sleeping disorders. These challenges negatively affect HCWs' professionalism, efficiency, and quality of life [2]. A recent study published by the World Economic Forum shows that approximately 50% of HCWs globally suffer from burnout, with this rate increasing to 66% among nurses and physicians [3]. These statistics underscore the urgent need to address the mental health crisis among HCWs to ensure quality care delivery.

Peer support programs (PSP), in which HCWs receive social support from peers, are a recognized type of intervention to mitigate burnout and improve well-being. Structured PSP with a goal-oriented approach, trained facilitators, and established curriculum in scheduled sessions have proven effective in preventing or mitigating burnout, anxiety, stress and depression among HCWs [4], [5].

These programs have been increasingly delivered through information and communication technologies (ICT) such as email, online surveys and tools, online forums, video conferencing services, social media, and text messaging applications [6]. In this research, we refer to this selection and use of ICT as peer support systems (PSS).

While PSS bring advantages such as increased accessibility and flexibility for delivering PSP, they also pose challenges related to participants' fear of sharing information and communication [7], perceived lack of safety and security [8], and low participation rate [9], [10]. These issues hinder the success of PSS and remain to be addressed to better enable their effectiveness.

We propose that these issues arise from a lack of trust in technology and other individuals when interactions are mediated by technology. Trust is essential in contexts perceived as risky by stakeholders, for example, when HCWs taking part in PSP discuss mental health issues subject to stigma that can affect their career and job security [11]. Previous research [12] has shown that trust fosters knowledge adoption and contribution, relational closeness, and technology usage intentions, hence behavioural trust outcomes that could address these issues and mitigate PSS-related challenges.

The notion of trust has been extensively studied in relationship to trustworthy systems. "Trustworthy" refers to properties of a system that satisfy non-functional requirements related to privacy, security, reliability, availability, safety, resilience and more [13], [14], so that stakeholders' trust towards the system may be warranted. However, trust and trustworthiness are not synonymous [14]. Trust is an attitude that a stakeholder holds towards a system and that may change over time (subjective), while trustworthiness is a system property (objective) [15]. Ideally, both concepts should be realized: a system should be trustworthy and actually trusted by stakeholders. However, the field of Requirements Engineering (RE) has mostly focused on trustworthiness because designers have more control over system features, and trustworthiness is perceived as having more sustainable outcomes compared to trust [15]. Moreover, trust is difficult to operationalize as requirements because it is subjective and perceptual [15], [16].

Nevertheless, understanding how to specify requirements related to perceptual trust remains necessary, since users may not perceive a trustworthy system as such. Imagine an individual who is reluctant to use an online banking application because they are concerned about privacy breaches, lack of understanding of how the technology works, and bad experiences they or their relatives may have had with technology in the past. Although the app meets trustworthy requirements such as security and privacy, the person does not trust this app. This lack of perceptual trust leads them to avoid using the app, despite its trustworthiness, and instead, they choose to visit the bank in person for their transactions [17]. Indeed, whether someone trusts a system depends not only on its design and the stakeholders' interaction with it [15], but also heavily on stakeholder properties such as their mindset, general propensity to trust technology, and prior knowledge and experience with similar systems [18].

We propose that gaining a deeper understanding of the concept of trust by revisiting extant theories of trust can lead to a more comprehensive identification of trust-related requirements. This approach is novel in that it allows us to systematically unpack trust into its varied types and antecedents and then elicit trust-related requirements based on these antecedents. Trust antecedents refer to the factors that precede or influence the development of trust [19]. Antecedents are a logical source of requirements, since they can be mapped to desired behavioral outcomes and represent factors that can be manipulated through system features. For instance, in designing a trust-based website for users, *interface design features* is defined as a trust antecedent and based on this antecedent, *simplicity* is one of the requirement, which relates to ease-of-use of navigation [20]. By adopting this approach, this research makes the following contributions:

- The paper proposes a trust framework that integrates several recognized theories of trust with varied foci (e.g., interpersonal trust, system trust) that is applicable to digital environments where trustees, parties which receive trust, are both individuals and technology. The framework identifies key types of trust in these contexts as well as their antecedents, hence the factors that can influence each type of trust. It thus provides a source of potential trust-related requirements for designing systems in which both "trust in technology" and "trust in others", mediated by technology, play a key role. This bridges the gap between perceptual trust and system requirements, extending the body of knowledge on trust-related requirements beyond those focused on trustworthiness as a systems property.

- By mapping the proposed trust framework to core PSS-supported activities, the paper shows a practical example of how the trust framework can be used to elicit trust-related requirements. This approach is also a first step towards designing a more effective PSS, a type of system that has received limited attention from the RE community [21].

- The paper also contributes to the existing body of knowledge on the use of theories from fields outside of RE – such as psychology and social sciences – by showing which components of such theories are amenable to the identification of requirements. We also outline a planned study for creating a prototype PSS using the framework and validating identified requirements, illustrating a path for moving from such theories to requirements to objectively measurable outcomes.

The rest of the paper is organized in the following manner. In section II, we provide the background on the concept of trust and explain why we have developed a trust framework for this study. Then, in section III, we present the trust framework and explain how we developed it. In section IV, we provide an illustrative case with four examples to demonstrate how the framework can be used to elicit requirements for PSS during a meeting. We present related work in section V, discuss the contributions of our study in section VI, and finally, describe our future research agenda in section VII.

## II. BACKGROUND

The concept of trust is multifaceted and significantly varies across disciplines. Trust can be dissected into various perspectives, including discipline-specific interpretations, dimensionality (uni- or multi-dimensional), and sources of trust (cognitive or affective). These perspectives shape the definition and understanding of trust, underlining its complexity. For instance, from a philosophical standpoint, trust can be understood as a trustor's (party who grants trust) voluntary vulnerability towards a trustee [22]. In contrast, the psychological perspective often views trust as a personality trait that influences individual behaviour and relationships from an early age, indicating a difference between trust as a choice and an innate disposition shaped by experiences [20]. From a sociological angle, trust can rather be conceptualized as a subjective probability affecting predictions about an agent's actions in social interactions [23]. Thus, despite the growing prominence of trust research, the literature remains varied in both conceptualization and measurement.

Despite these variations, a thorough examination of trust theories, models, and frameworks shows that they mostly focus on three main constructs: "trust antecedents", "trust types" and "trust outcomes". For instance, seminal literature has explored trust antecedents and outcomes within organizations [24]. Other authors have instead differentiated between external and internal trust antecedents without specifying trust types or outcomes [25]. More recent literature has explored the relationship between trust and technology, for example, by proposing a detailed trust framework for health online communities that identifies trust antecedents such as *information credibility* and *community support*; behavioural trust outcomes such as *knowledge adoption* and *engagement*; and relationships among these elements [12].

While a thorough literature review on trust theories is beyond the scope of this paper, this brief outline shows that trust is a multi-dimensional concept that cannot be truly understood by selectively focusing on a single theory or model of trust. In the context of PSS, for example, HCWs must trust both the technology and the mediated interactions that they will have with their peers. As such, diverse types of trust and trust antecedents should be considered to generate a comprehensive set of requirements for these systems.

## III. A Trust Framework for PSS

We propose a trust framework that presents a broad set of *trust antecedents*, *trust types*, *trust outcomes* and interrelationships among these constructs. The framework is anchored in recognized trust theories from different disciplines such as computer science and information systems (e.g., [12], [18], [26]), and management and social sciences (e.g., [27], [28], [29]).

While the proposed framework aligns with the foundational constructs of trust identified in the literature, it extends beyond them by integrating all constructs and their categories in a unified framework. A core model of trust integrated into the framework is a model for trust in specific technology such as Excel, which identifies *trust antecedents* (e.g., propensity to trust in general technology) leading to *trust in technology* and resulting *outcomes* (e.g., intention to explore and deep structure use) [18]. However, this seminal study does not address trust in group contexts. A trust model tailored to virtual communities for patients was thus integrated [26]. It presents *trust antecedents* (perceived competence and perceived goodwill) that affect three *types of trust (dispositional trust, interpersonal trust and system trust)* that are important in this context. Despite addressing different *types of trust* and *trust antecedents*, this study did not delve into *trust outcomes* or provide detailed insights into the *trust antecedents*.

Complementary models of trust were sought to ensure the comprehensiveness of the framework. For example, [28] presents a multi-level model of trust in teams, emphasizing three primary constructs of trust, along with their categories, sub-categories and their interrelationships: *trust antecedents* (team-level factors and individual-level factors); *types of trust* (team trust and interpersonal trust between members); and, *trust outcomes* (team-level outcomes and individual-level outcomes). Similarly, [27] presented a trust model encompassing *trust antecedents* (benevolence, ability, integrity and propensity to trust), *trust* (as a unique construct) and *outcomes* and their linear relationships, with a focus on organizational context.

By integrating several theories from different disciplines, this framework provides a deep understanding of the relevant trust dynamics for digitally-enabled contexts such as PSS. The framework, shown in Figure 1, identifies and relates key *trust antecedents*, *trust types*, and *trust outcomes*. Each of these constructs is further classified, providing detailed categories under each main construct.

In the framework, dashed boxes, such as the one surrounding *dispositional trust* and *organizational factors*, indicate trust constructs recognized in the literature but not examined in this study because they are not amenable to identifying software requirements. For example, while *trust antecedents* related to *organizational factors*, such as *human resource management practices*, can have an important impact on HCWs' trust toward PSS, they cannot be controlled or influenced through PSS. Similarly, dashed lines suggest relationships that have been identified in literature but that have not received significant support, while solid lines depict established relationships. Arrows indicate the directionality of relationships, either unidirectional or bidirectional.

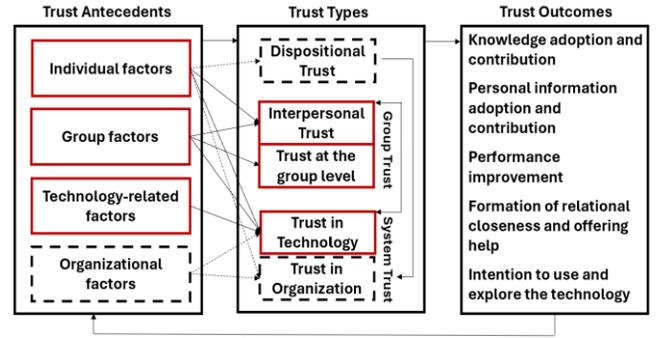

Fig. 1. Integrated Trust Framework for Digitally-enabled Contexts

The Integrated Trust Framework for Digitally-enabled Contexts contains three types of trust: *dispositional trust*, *group trust*, and *system trust*. *Dispositional trust* refers to each healthcare worker's inherent tendency to trust trustees such as peers, organizations [24], [27] and technology [18]. *Group trust* refers to a) interpersonal trust among HCWs at the individual level and b) trust collectively shared among group members [28]. *System trust* refers to trust in organizational entities [27], like hospitals, and trust in specific technology [18], such as PSS. Table I unpacks the categories of trust antecedents that are relevant for PSS, identifying their sub-categories and antecedents as identified in the literature. To make it easier to follow the table, we have assigned numbers to each item.

TABLE I. TRUST ANTECEDENTS

| Categories of Trust antecedents | Sub-categories | Antecedents |
|---|---|---|
| *1. Individual factors* | 1.1. Trustor characteristics | 1.1.1. Propensity to trust in people and technology<br>1.1.2. Life experiences, personality types, cultural backgrounds, education, and several other socioeconomic factors |
| | 1.2. Trustee characteristics | 1.2.1. Trustee's perceived ability, benevolence, and integrity<br>1.2.2. Trustee reputation |
| | 1.3. Interpersonal relationships | 1.3.1. Both parties' history<br>1.3.2. Degree of similarity in common goals and values and occupation |
| *2. Group factors* | ------------ | 2.1.1. Group composition<br>2.1.2. Relationship structure<br>2.1.3. Group climate |
| *3. Technology-related factors* | 3.1. Technology structure factors | 3.1.1. Structural assurance<br>3.1.2. Trustworthy system features |
| | 3.2. Human-technology interaction factors | 3.2.1. Situational normality<br>3.2.2. System interface features<br>3.2.3. Reputation of the platform<br>3.2.4. Perceived usefulness<br>3.2.5. Perceived ease of use |

Trust outcomes delve into the repercussions of established trust, encompassing both the behaviours and intentions of the trustor and the broader impacts on trustees. While behavioural trust outcomes such as *information contribution* [12] and *team collaboration* [28] are more tangible, trust intentions such as *the intention to explore the technology* [30] are more subjective, reflecting the trustor's willingness to depend on the trustee and their perception of the trustee's trustworthiness [31].

To identify trust-related requirements for PSS, we map *trust antecedents* to activities supported by PSS (TABLE II). The first column of Table II lists the core activities of structured PSP that should be supported by a PSS: *recruiting*, *scheduling*, *informing*, *reminding*, *meetings*, *communicating*, and *getting feedback* [8]. The column also includes common tools used in each activity. The second column of the table lists different *types of trust* that are relevant for each type of activity. This is followed by related categories of *trust antecedents* as potential sources of requirements for each activity in the table's last three columns.

In constructing this table, the nature of each activity thus dictates the *types of trust* involved. For example, activities centred on peer interaction relate to *group trust*, while other activities may relate solely to *trust in technology*. By examining the relationship between *types of trust* and *trust antecedents*, we can then map relevant *trust antecedents* to each activity. For example, in the case of a "Meeting" activity, both *group trust* and *trust in technology* play significant roles. Therefore, *trust antecedents* that influence this activity include *individual factors*, *group factors*, and *technology-related factors* (see Fig. 1). Such mapping can be used to guide elicitation activities by identifying relevant concerns to explore with stakeholders in relation to core activities supported by a system. We explain the application of the framework in further detail in Section IV.

TABLE II. MAPPING TRUST ANTECEDENTS TO PSS

| PSS-Supported Activities (and Common Tools) | Trust Types | Trust Antecedents | | |
|---|---|---|---|---|
| | | *Individual factors* | *Technology related factors* | *Group factors* |
| **Recruiting** (email) | -Trust in technology | | ✓ | |
| **Scheduling** (online scheduling) | -Trust in technology | | ✓ | |
| **Informing** (email) | -Trust in technology | | ✓ | |
| **Reminding** (email) | -Trust in technology | | ✓ | |
| **Meeting** (videoconferencing) | -Group Trust -Trust in technology | ✓ | ✓ | ✓ |
| **Communicating** (text messaging, online forums) | -Group Trust -Trust in technology | ✓ | ✓ | ✓ |
| **Getting feedback** (online surveys) | -Trust in technology | ✓ | ✓ | |

## IV. ILLUSTRATIVE SCENARIO

In this section, we explain how the Integrated Trust Framework can be used to elicit requirements through its application to an illustrative scenario of HCWs and peer support facilitators using a PSS for a peer support program meeting. As shown in Table II, three categories of *trust antecedents* may be relevant in a PSS-supported "Meeting" activity: *individual factors*, *technology-related factors* and *group factors*. The antecedents related to these categories, shown in Table I, should thus be used as a source of questions for PSS users, including HCWs and peer support program facilitators. We assume that certain antecedents will be emphasized by users, while some may be found to be irrelevant from their perspectives.

We provide four examples to explain the process of eliciting trust-related requirements using the Integrated framework and offer specific requirements as examples.

### A. Scenario

The PSS-supported activity "Meeting" enables HCWs to gather virtually at scheduled times to discuss specific topics related to well-being and mental health. A trained peer support program facilitator, who is also a peer, leads each session, creating a supportive environment where HCWs can share experiences and coping mechanisms while offering mutual support [8]. This fosters a sense of community among HCWs and improves their mental health and resilience.

### B. Eliciting Trust-related Requirements

Both *group trust* and *trust in technology* are important in the meeting activity. A healthcare worker should trust their peers (e.g., John, who is a physician) and the technology used for the meeting (e.g., Zoom). *Group trust* allows effective communication and collaboration, and *trust in technology* facilitates communication and information exchange.

By applying our framework, we know that three categories of *trust antecedents* need to be considered: *individual factors*, *group factors*, and *technology-related factors*. We use the list of *trust antecedents* provided in Table I to guide our questions to stakeholders and elicit trust-related requirements for PSS-supported meetings.

We provide four examples of trust types and antecedents relevant to this activity, along with high-level system requirements they could generate. To follow the process of eliciting trust-related requirements, we refer to numbers in Table 1.

**Example 1: Group Composition (2.1.1).** *Group trust* is important for interpersonal communication in peer support sessions. Group composition (2.1.1) is an antecedent that influences *group trust* [58]. Understanding what characteristics of individuals in a group influence group trust is crucial, as studies in PSP have found that HCWs are not willing to share personal information in groups with mixed hierarchical levels. Both higher and lower levels of the hierarchy experience this issue. Nurses report that the presence of senior staff during sessions makes their tone more formal, leading them to withhold personal information [32].

Similarly, senior staff also tend to censor their input due to the hierarchical relationships within the group [33].

Therefore, elicitation questions focused on users' needs regarding group composition should be asked, such as: "How could members of a group within a peer support session be selected for you to feel comfortable sharing personal information?" Answers to this question could be more nuanced than what is suggested in literature. For example, some stakeholders could prefer to include participants from different hierarchical levels in their sessions to ensure inclusiveness, while others favour participants from the same hierarchical position to have more similarity. To address this variation, trust-related requirements for a PSS could be that the system "asks users about their preferences for group composition (similar level vs. varied levels)" and "automatically match users based on their hierarchical position".

In this example, requirements elicitation questions focused on Group composition (2.1.1), an antecedent in *group factors* (2), leads to the identification of two trust-related system requirements for PSS.

**Example 2: Situational normality in technology (3.2.1)**

*Trust in technology* is another important *type of trust* in peer support sessions. Situational normality in technology (3.2.1.) is an antecedent that influences *technology in trust* [30]. This term embodies users' perceptions of the technology's benevolence, competence, and integrity [30], [31]. Perceived competence refers to the belief of users regarding the capability and functionality of a system [30].

Therefore, elicitation questions focused on users' beliefs regarding the competence of a system should be asked. For example: "In your experience using PSS, are there specific features or aspects of the system's functionality that contribute to or detract from your perception of its competence? How do you think the system could enhance its perceived competence to better meet your needs or expectations?". Answers to these questions could focus on data privacy, which is a known concern in the context of a health system [26]. A study highlighting the reluctance among HCWs to participate in a WeChat peer support group in China attributed their hesitancy to fears of government scrutiny and political repercussions [10], which can arise from a lack of transparency about the use and protection of data. It is crucial for HCWs to have clear and accessible information about if, how, and where their information might be collected, used, shared, and retained.

In this regard, users could suggest that PSS should provide a clear and easily accessible privacy policy that explains how user data is collected, used, stored, and protected. This could include detailing any data encryption methods, access controls, data retention policies, information about who has the authority to access their information and the specific conditions for granting access and making that information accessible before and during the meeting activity. Related trust-related requirements for PSS could be that the system should provide "transparency of use, protection and retention of data" and "transparency of access rights and authorization levels in the system".

This example focuses on *trust in technology,* influenced by *technology-related factors* (3). Situational normality in technology (3.2.1), as a key antecedent in this category, leads to the identification of two trust-related system requirements for PSS.

**Example 3: Perceived usefulness (3.2.4)**

Perceived usefulness (3.2.4) is another antecedent that influences *technology in trust* [25]. This concept refers to how well the technology meets users' needs and helps them achieve their goals. For example, in the context of PSS for HCWs, usefulness encompasses factors such as how effectively it facilitates communication and fosters collaboration.

Therefore, elicitation questions focused on users' beliefs regarding the usefulness of a system should be asked. For example, "In your opinion, what features or functionalities would make a peer support system most beneficial and effective for you and your colleagues? How do you think a peer support system could contribute to improving teamwork and communication during peer support meetings?". These questions could be answered in terms of improving social presence, the sense of being together with others in digital environments, which can enhance the perceived usefulness of the system [34]. Social presence can enrich the user experience in PSS by creating a sense of community and engagement.

In this regard, HCWs might mention some examples related to social presence in PSS. HCWs might need to use feedback mechanisms such as likes or dislikes or emojis to enhance social presence in PSS to express emotions non-verbally. Also, they might express the desire to send a direct message to the facilitator or other peers to discuss an issue privately. Peer support facilitators might need to have collaborative tools such as whiteboards for group activities and ask HCWs to share their opinions on specific topics related to the session. Therefore, trust-related requirements for PSS could be that the system should "incorporate an interactive feedback mechanism", "possess collaborative tools", and "have a direct messaging feature".

In this example, elicitation questions related to Perceived usefulness (3.2.4), a key antecedent in *technology-related factors*, leads to the identification of three trust-related requirements for PSS.

**Example 4: Trustworthy system features (3.1.2)**

Trustworthy system features (3.1.2) is another antecedent that influences *technology in trust*. It encompasses various non-functional requirements of PSS such as privacy and security [14] and how they should be addressed.

Therefore, elicitation questions focused on users' beliefs regarding trustworthy features of a system should be asked. We assume that safeguarding privacy is essential for them, because HCWs exchange personal and sensitive information in PSS. In this regard, some questions we could ask are "What specific privacy controls or features would you like to see implemented in this system to safeguard privacy? How important is it for you to maintain the privacy of your identity when evaluating a system like this?".

These questions could be addressed by a privacy policy regarding data access and management, also mentioned in Example 2. This is to be expected since the same requirement could address more than one user need. Stakeholders' responses to stated questions could also relate to a need for anonymity during meetings to protect their identity, perhaps suggesting the use of customizable avatars as a digital representation to maintain user anonymity while allowing for the expression of personal traits and emotions. Such answers would be in line with evidence that avatar usage, as seen on platforms like Second Life, can mitigate communication barriers in PSS [35]. Accordingly, PSS could provide the functionality for HCWs to participate in sessions either as themselves or through an avatar. The resulting trust-related requirements for PSS would be that the system should support "identity management" and "anonymity through customizable avatar," which ensures users can control their identity and visibility, thus addressing their privacy concerns.

This example focuses on *trust in technology*, influenced by *technology-related factors* (3). Privacy as an important trustworthy system feature (3.1.2) leads to the identification of two trust-related system requirements for PSS.

## V. RELATED WORK

Trust is a fundamental concept studied across various relational contexts including interpersonal, organizational, and those involving technology. In the field of Requirements and Software Engineering, trust is approached in several ways.

One approach is to study trust as a measure to evaluate existing systems or trust states among people in a technology-related community. For instance, [36] propose a metric framework called TRAM (trust, resilience, and agility metrics) to measure trustworthy systems' quality. In this framework, "trust" focuses on measuring the security and performance aspects of a system as well as its predictability. Meanwhile, [37] decompose trust to measure it within open-source software communities. They decompose trust into trust components (e.g., benevolence, integrity) and trust contracts, representing the trustor's belief in the trustee's ability to complete a specific action. However, while metrics are essential for assessing trust, they do not directly lead to trust-related requirements for systems.

Other studies in this field have focused on providing a better understanding of the concept of trust with varied scope and complexity. For example, [38] present the mental aspects of trust and distinguishes between two types of trust (interpersonal and institution-based trust) by presenting behavioural aspects and risk emergence. [39] present a trust model in the context of a multimedia service environment, distinguishing trust relationships among different trustees (e.g., multimedia generator) and trustors (e.g., multimedia service provider) and their attributes. Additionally, [26] propose a trust model for virtual communities that includes three types of trust (dispositional trust, system trust and interpersonal trust) that play a role in this context, as well as trust antecedents that affect those types of trust. They also provide requirements (e.g., anonymity for users) related to trust antecedent (e.g., perceived goodwill) to implement for these virtual communities. Although these approaches provide a deeper understanding of trust, they either do not lead to the identification of requirements, or are not as comprehensive as the Integrated Trust Framework proposed in this research.

A significant body of research focuses on trustworthy systems, delineating attributes, and requirements. For instance, [13] discuss engineering trustworthy software systems for the US Air Force, emphasizing reliability, usability, interoperability, and security. [14] distinguishes between "trustworthy" and "trust," providing a trust model for analyzing trust formation between trustor and trustee, and identifying six attributes (e.g., security and reliability) and seven requirements essential for developing trustworthy systems. Similarly, [15] emphasizes the significance of trust and trustworthiness in developing and deploying AI systems, focusing on trustworthy features that lead to controllable requirements. Although these studies mostly focus on trustworthy features of the system and provide requirements for those, they do not focus on requirements related to perceptual trust.

## VI. DISCUSSION

The approach proposed in this research aims to extend our understanding of existing trust-related requirements by focusing on how users may perceive trust, in addition to conventional requirements for a trustworthy system [15], [36], [40]. We present four examples related to PSS-supported meetings. Three of these examples identify potential trust-related requirements that go beyond typical trustworthy system requirements. In our first example, the system should collect users' preferences for "Group composition" and match users based on their hierarchical positions. This focuses on the *group factors* category of *trust antecedent* that is known to enhance *group trust*, thus addressing the fear of sharing information and communicating, a recognized challenge faced by HCWs when using PSS [9]. The second and third examples aim to enhance *trust in technology*, which addresses HCWs' perceived lack of safety when using PSS [9]. Resulting examples of requirements concern "transparency in data usage, protection, and retention", as well as "transparency of data access rights and authorization levels" to enhance users' perceptions of the system's competence. In addition, the "perceived usefulness of a system" as an antecedent could lead to essential requirements such as "interactive feedback mechanisms", "collaborative tools", and "direct messaging features". These three examples illustrate requirements related to perceptual trust, which delve into what makes a system trustable from a user perspective, extending beyond specific system properties. The fourth example addresses a common feature of trustworthy systems – privacy – by focusing on the specific needs of HCWs through identity management and anonymity features. This example shows that the Integrated Trust Framework encompasses concepts of trustworthiness. The approach proposed in this paper thus provides a strong foundation for expanding our understanding of trust-related requirements while recognizing the importance of trustworthy system requirements.

Furthermore, this paper provides a foundation for future research on PSS for HCWs. While existing literature highlights the significance of PSP in addressing well-being issues among HCWs [11], [33], the systems used to deliver these programs are limited in their ability to do so effectively [9]. This issue has received limited attention from the RE community, despite some interesting propositions such as a peer-sourcing software architecture for peer support in mental health [21]. In addition, by focusing on enhancing trust within PSS, we aim to improve their functionality, ensuring they better meet the needs of HCWs and other professional groups who suffer from mental health issues and could benefit from PSP.

The Integrated Trust Framework and Trust antecedents table (TABLE I) can be valuable resources for identifying potential requirements for systems that rely heavily on trust both in technology itself, and in technology as a mediator of interactions between users. Besides PSS for HCWs, which was the main focus of this paper, the framework and trust antecedents table can be applied to PSS in other contexts and for other users, such as people who have experienced cancer [41] or virtual health communities[12], [26]. In both contexts, *interpersonal trust* (trust between users) and *trust in the technology* play crucial roles. The developed framework explicitly addresses both *types of trust* and their underlying *antecedents*. For instance, in virtual health communities, where individuals seek support and information, trust in both fellow community members and the technology platform facilitating interactions is essential for fostering engagement and well-being [26]. Therefore, the Integrated Trust Framework and its associated trust antecedents are highly applicable to these contexts.

Finally, our study contributes to the discussion on the use of theories from external fields in RE. A significant body of work has shown that such theories can be used to identify new and more comprehensive system requirements [42] beyond those that can be elicited solely from individuals or groups [43]. A subset of this work focuses more specifically on articulating human and social behaviours by leveraging extant theory, for example, by operationalizing the concept of trust as an ontology [38]. Such research recognizes the need for a more in-depth understanding of human and social behaviours in contexts where fears and risk aversions could deter system use, such as health care. However, there is typically a large gap between such in-depth understandings and system requirements. This research shows that focusing on the variables that influence human behaviours – antecedents – provides a basis for investigating how system features can leverage these influences to achieve desired outcomes. As such, it provides a novel yet practical approach for mapping extant theories to requirements elicitation activities where context-relevant antecedents can be identified and mapped to system requirements.

## VII. CONCLUSION AND FUTURE WORK

This paper proposes an Integrated Trust Framework for Digitally-enabled Contexts where trust in technology and trust in people mediated by technology is crucial. The developed framework integrates various trust theories from different fields, such as information systems and computer science (e.g., [20], [30], [31]), organizational behaviour and human resources (e.g., [27], [28]), and psychology (e.g., [19], [22]) to account for the multi-dimensional nature of trust. The integrated framework has three core constructs: *trust antecedents*, *trust types*, and *trust outcomes*, and explains their interrelations. It classifies trust into *dispositional trust*, *group trust*, and *system trust* and identifies different *trust antecedents* that influence each type of trust. Doing this demonstrates how *trust antecedents* impact different *types of trust*, which, in turn, lead to specific behavioural and intentional *trust outcomes*.

Importantly, the Integrated Trust Framework shows how the understanding of *trust antecedents* can be used to identify system requirements that can help address known challenges related to the use of PSS by HCWs, thus enhancing the effectiveness of these systems. Moreover, the proposed framework provides a comprehensive understanding of the notion of trust, unpacking what trust means from the perspective of system users and how they may perceive a system to be worthy of their trust beyond system properties typically associated with trustworthiness. The novelty of this work thus stems from its focus on perceptual trust and from its demonstration of how perceptual trust can be mapped to systems requirements through the understanding of trust antecedents. Despite its contributions, this work shows limitations. Particularly because the focus of this work is on PSS, antecedents related to "trust in organizations" were not considered as potential sources of requirements. However, they may be relevant for other types of systems in which trust plays a key role, such as enterprise systems or systems focused on human resource management.

The next step of this research will focus on getting expert input on the framework as an initial validation of its constructs and their relevance. The refined framework will then be used to guide requirements elicitation during a study in which a prototype PSS will be developed with HCWs. The iterative nature of the study will allow for refining the framework itself for the specific context of PSS for HCWs through the identification of the most relevant antecedents for this context. The results of the study will also include a mapping of these antecedents to high-level requirements and to their realization in the prototype. We plan to share these results with the RE community alongside methodological guidance for eliciting trust requirements, for example, through requirements elicitation questions and a glossary of each concept identified in Table I, in the form of open-source tools made publicly available on GitHub.

We believe that this research provides the foundations for a broader research program on trust-related requirements for a variety of systems in digitally-enabled contexts. Future research could, for example, focus on identifying more comprehensively which type of trust is important for which type of system, such as through a taxonomy of trust-centric systems. For example, dispositional trust is likely to play a key role in contexts where individual users are asked to use a novel technology such as social robots, while it may be of less consequence in contexts where users use a system to accomplish tasks with a commonly-used technology. Also, tools to facilitate stakeholders' ranking of the importance of varied trust antecedents in their context of use could be developed to increase efficiency during trust-related requirements elicitation activities.

ACKNOWLEDGMENT

This research is partially funded by an NSERC Discovery Grant. We would like to acknowledge the valuable intellectual contributions of Dr. Daniel Amyot and Dr. Jennifer Dimoff from the University of Ottawa.

REFERENCES

[1] A. E. Muller *et al.*, "The mental health impact of the covid-19 pandemic on healthcare workers, and interventions to help them: A rapid systematic review," *Psychiatry Research*, vol. 293, p. 113441, Nov. 2020, doi: 10.1016/j.psychres.2020.113441.

[2] L. E. Søvold *et al.*, "Prioritizing the Mental Health and Well-Being of Healthcare Workers: An Urgent Global Public Health Priority," *Front. Public Health*, vol. 9, p. 679397, May 2021, doi: 10.3389/fpubh.2021.679397.

[3] D. Agenda, "Why addressing burnout among healthcare workers is crucial to advance health-related SDGs," World Economic Forum. [Online]. Available: https://www.weforum.org/agenda/2023/09/addressing-healthcare-worker-burnout-and-the-urgent-path-to-sdg3-health/

[4] U. Peterson, G. Bergström, M. Samuelsson, M. Åsberg, and Å. Nygren, "Reflecting peer-support groups in the prevention of stress and burnout: randomized controlled trial," *Journal of Advanced Nursing*, vol. 63, no. 5, pp. 506–516, Sep. 2008, doi: 10.1111/j.1365-2648.2008.04743.x.

[5] E. A. Keyser, L. F. Weir, M. M. Valdez, J. K. Aden, and R. I. Matos, "Extending Peer Support Across the Military Health System to Decrease Clinician Burnout," *Military Medicine*, vol. 186, no. Supplement_1, pp. 153–159, Jan. 2021, doi: 10.1093/milmed/usaa225.

[6] R. Carbone *et al.*, "Peer support between healthcare workers in hospital and out-of-hospital settings: a scoping review," *Acta Biomedica Atenei Parmensis*, vol. 93, no. 5, p. e2022308, Oct. 2022, doi: 10.23750/abm.v93i5.13729.

[7] C. Jenkins, J. Oyebode, S. Bicknell, N. Webster, P. Bentham, and A. Smythe, "Exploring newly qualified nurses' experiences of support and perceptions of peer support online: A qualitative study," *Journal of Clinical Nursing*, vol. 30, no. 19–20, pp. 2924–2934, Oct. 2021, doi: 10.1111/jocn.15798.

[8] S. Samnani and M. Awal, "Balint in the time of COVID-19: Participant and facilitator experience of virtual Balint groups compared with in-person," *Int J Psychiatry Med*, p. 0091217421105373, Feb. 2022, doi: 10.1177/00912174211053733.

[9] Y. Gheidar, L. Lessard, and Y. Yao, "Integrating the Voice of Healthcare Workers in Requirements Elicitation: A Balance Between Rigour and Relevance," in *2023 IEEE 31st International Requirements Engineering Conference Workshops (REW)*, Hannover, Germany: IEEE, Sep. 2023, pp. 446–451. doi: 10.1109/REW57809.2023.00084.

[10] P. Cheng *et al.*, "COVID-19 Epidemic Peer Support and Crisis Intervention Via Social Media," *Community Ment Health J*, vol. 56, no. 5, pp. 786–792, Jul. 2020, doi: 10.1007/s10597-020-00624-5.

[11] M. B. Korman *et al.*, "Implementing the STEADY Wellness Program to Support Healthcare Workers throughout the COVID-19 Pandemic," *Healthcare*, vol. 10, no. 10, p. 1830, Sep. 2022, doi: 10.3390/healthcare10101830.

[12] R. Connolly *et al.*, "Understanding Engagement in Online Health Communities: A Trust-Based Perspective," *JAIS*, vol. 24, no. 2, pp. 345–378, 2023, doi: 10.17705/1jais.00785.

[13] N. Subramanian, S. Drager, and W. McKeever, "Designing Trustworthy Software Systems Using the NFR Approach," in *Emerging Trends in ICT Security*, Elsevier, 2014, pp. 203–225. doi: 10.1016/B978-0-12-411474-6.00013-X.

[14] J. Li, B. Mao, Z. Liang, Z. Zhang, Q. Lin, and X. Yao, "Trust and Trustworthiness: What They Are and How to Achieve Them," in *2021 IEEE International Conference on Pervasive Computing and Communications Workshops and other Affiliated Events (PerCom Workshops)*, Kassel, Germany: IEEE, Mar. 2021, pp. 711–717. doi: 10.1109/PerComWorkshops51409.2021.9430929.

[15] L. Kastner, M. Langer, V. Lazar, A. Schomacker, T. Speith, and S. Sterz, "On the Relation of Trust and Explainability: Why to Engineer for Trustworthiness," in *2021 IEEE 29th International Requirements Engineering Conference Workshops (REW)*, Notre Dame, IN, USA: IEEE, Sep. 2021, pp. 169–175. doi: 10.1109/REW53955.2021.00031.

[16] S. Amershi *et al.*, "Guidelines for Human-AI Interaction," in *Proceedings of the 2019 CHI Conference on Human Factors in Computing Systems*, Glasgow Scotland Uk: ACM, May 2019, pp. 1–13. doi: 10.1145/3290605.3300233.

[17] M. Minnes and T. Saskin, "Helping seniors navigate the financial marketplace through the use of financial technology," University of Waterloo, Nov. 2021. [Online]. Available: https://www.canada.ca/en/financial-consumer-agency/programs/research/2021-building-better-financial-futures-challenge/helping-seniors-navigate-financial-marketplace.html

[18] D. H. Mcknight, M. Carter, J. B. Thatcher, and P. F. Clay, "Trust in a specific technology: An investigation of its components and measures," *ACM Trans. Manage. Inf. Syst.*, vol. 2, no. 2, pp. 1–25, Jun. 2011, doi: 10.1145/1985347.1985353.

[19] H. Gill, K. Boies, J. E. Finegan, and J. McNally, "Antecedents Of Trust: Establishing A Boundary Condition For The Relation Between Propensity To Trust And Intention To Trust," *J Bus Psychol*, vol. 19, no. 3, pp. 287–302, Apr. 2005, doi: 10.1007/s10869-004-2229-8.

[20] Y. D. Wang and H. H. Emurian, "An overview of online trust: Concepts, elements, and implications," *Computers in Human Behavior*, vol. 21, no. 1, pp. 105–125, Jan. 2005, doi: 10.1016/j.chb.2003.11.008.

[21] M. Honary, J. Lee, C. Bull, J. Wang, and S. Helal, "What Happens in Peer-Support, Stays in Peer-Support: Software Architecture for Peer-Sourcing in Mental Health," in *2020 IEEE 44th Annual Computers, Software, and Applications Conference (COMPSAC)*, Madrid, Spain: IEEE, Jul. 2020, pp. 644–653. doi: 10.1109/COMPSAC48688.2020.0-184.

[22] A. Baier, "Trust and Antitrust," *Ethics*, vol. 96, no. 2, pp. 231–260, Jan. 1986, doi: 10.1086/292745.

[23] D. Gambetta, "Can We Trust Trust?," in *Trust: Making and Breaking Cooperative Relations*, D. Gambetta, Ed., Blackwell, 1988, pp. 213–237.

[24] D. H. McKnight, L. L. Cummings, and N. L. Chervany, "Initial Trust Formation in New Organizational Relationships," *The Academy of Management Review*, vol. 23, no. 3, p. 473, Jul. 1998, doi: 10.2307/259290.

[25] J. Salo and H. Karjaluoto, "A conceptual model of trust in the online environment," *Online Information Review*, vol. 31, no. 5, pp. 604–621, Oct. 2007, doi: 10.1108/14684520710832324.

[26] J. M. Leimeister, W. Ebner, and H. Krcmar, "Design, Implementation, and Evaluation of Trust-Supporting Components in Virtual Communities for Patients," *Journal of Management Information Systems*, vol. 21, no. 4, pp. 101–131, Apr. 2005, doi: 10.1080/07421222.2005.11045825.

[27] R. C. Mayer, J. H. Davis, and F. D. Schoorman, "An Integrative Model of Organizational Trust," *The Academy of Management Review*, vol. 20, no. 3, p. 709, Jul. 1995, doi: 10.2307/258792.

[28] A. C. Costa, C. A. Fulmer, and N. R. Anderson, "Trust in work teams: An integrative review, multilevel model, and future directions," *J Organ Behav*, vol. 39, no. 2, pp. 169–184, Feb. 2018, doi: 10.1002/job.2213.

[29] M. Ter Huurne, A. Ronteltap, R. Corten, and V. Buskens, "Antecedents of trust in the sharing economy: A systematic review," *J Consumer Behav*, vol. 16, no. 6, pp. 485–498, Nov. 2017, doi: 10.1002/cb.1667.

[30] D. H. Mcknight, M. Carter, J. B. Thatcher, and P. F. Clay, "Trust in a specific technology: An investigation of its components and measures," *ACM Trans. Manage. Inf. Syst.*, vol. 2, no. 2, pp. 1–25, Jun. 2011, doi: 10.1145/1985347.1985353.

[31] X. Li, T. J. Hess, and J. S. Valacich, "Why do we trust new technology? A study of initial trust formation with organizational information systems," *The Journal of Strategic Information Systems*, vol. 17, no. 1, pp. 39–71, Mar. 2008, doi: 10.1016/j.jsis.2008.01.001.

[32] N. Webster, J. Oyebode, C. Jenkins, and A. Smythe, "Using technology to support the social and emotional well-being of nurses: A scoping review protocol," *J Adv Nurs*, vol. 75, no. 4, pp. 898–904, Apr. 2019, doi: 10.1111/jan.13942.

[33] C. A. Connors, V. Dukhanin, A. L. March, J. A. Parks, M. Norvell, and A. W. Wu, "Peer support for nurses as second victims: Resilience, burnout, and job satisfaction," *Journal of Patient Safety and Risk Management*, vol. 25, no. 1, pp. 22–28, Feb. 2020, doi: 10.1177/2516043519882517.

[34] N. Lankton, D. H. McKnight, and J. Tripp, "Technology, Humanness, and Trust: Rethinking Trust in Technology," *JAIS*, vol. 16, no. 10, pp. 880–918, Oct. 2015, doi: 10.17705/1jais.00411.

[35] K. L. Rice, M. J. Bennett, and L. Billingsley, "Using Second Life to Facilitate Peer Storytelling for Grieving Oncology Nurses," vol. 14, no. 4, p. 12, 2014.

[36] J.-H. Cho, P. M. Hurley, and S. Xu, "Metrics and measurement of trustworthy systems," in *MILCOM 2016 - 2016 IEEE Military Communications Conference*, Baltimore, MD, USA: IEEE, Nov. 2016, pp. 1237–1242. doi: 10.1109/MILCOM.2016.7795500.

[37] L. Boughton, C. Miller, Y. Acar, D. Wermke, and C. Kästner, "Decomposing and Measuring Trust in Open-Source Software Supply Chains," in *IEEE/ACM 46th International Conference on Software Engineering: New Ideas and Emerging Results (ICSE-NIER '24)*, IEEE/ACM, Apr. 2024.

[38] G. Amaral, T. P. Sales, G. Guizzardi, and D. Porello, "Towards a Reference Ontology of Trust," in *On the Move to Meaningful Internet Systems: OTM 2019 Conferences*, vol. 11877, H. Panetto, C. Debruyne, M. Hepp, D. Lewis, C. A. Ardagna, and R. Meersman, Eds., in Lecture Notes in Computer Science, vol. 11877. , Cham: Springer International Publishing, 2019, pp. 3–21. doi: 10.1007/978-3-030-33246-4_1.

[39] S. Moon, S. Jung, and S. Jung, "A study on trust factors in a multimedia service environment," in *2018 International Conference on Information Networking (ICOIN)*, Chiang Mai: IEEE, Jan. 2018, pp. 67–69. doi: 10.1109/ICOIN.2018.8343086.

[40] G. Heiser, T. Murray, and G. Klein, "It's Time for Trustworthy Systems," *IEEE Secur. Privacy Mag.*, vol. 10, no. 2, pp. 67–70, Mar. 2012, doi: 10.1109/MSP.2012.41.

[41] M. Jablotschkin, L. Binkowski, R. Markovits Hoopii, and J. Weis, "Benefits and challenges of cancer peer support groups: A systematic review of qualitative studies," *European J Cancer Care*, vol. 31, no. 6, Nov. 2022, doi: 10.1111/ecc.13700.

[42] L. Lessard, D. Amyot, O. Aswad, and A. Mouttham, "Expanding the nature and scope of requirements for service systems through Service-Dominant Logic: the case of a telemonitoring service," *Requirements Eng*, vol. 25, no. 3, pp. 273–293, Sep. 2020, doi: 10.1007/s00766-019-00322-z.

[43] R. Chitchyan and C. Bird, "Theory as a source of software and system requirements," *Requirements Eng*, vol. 27, no. 3, pp. 375–398, Sep. 2022, doi: 10.1007/s00766-022-00380-w.